%% file: preprint.tex
\documentclass[aps,prd,amsfonts,amssymb,amsmath,nofootinbib,notitlepage, 12pt,tightenlines,longbibliography]{revtex4-1}
\usepackage{breqn, esint, graphics, hyperref, array, xcolor, mathtools, longtable}

\newcolumntype{L}[1]{>{\raggedright\let\newline\\\arraybackslash\hspace{0pt}}m{#1}}
\newcolumntype{C}[1]{>{\centering\let\newline\\\arraybackslash\hspace{0pt}}m{#1}}
\newcolumntype{R}[1]{>{\raggedleft\let\newline\\\arraybackslash\hspace{0pt}}m{#1}}

\newcommand{\RN}[1]{%
  \textup{\uppercase\expandafter{\romannumeral#1}}%
}

\begin{document}

\title{Hagedorn Temperature in Superstring Bits and SU(N) Characters}

\author{Sourav Raha}
\email[Email: ]{souravraha@ufl.edu}
%\homepage[Visit: ]{}
\affiliation{Department of Physics$,$  University of Florida}
\date{\today}
\begin{abstract}
We study the simplest superstring bit model at finite $N$ using the
characters of the $ SU (N)$ group. We obtain exact, analytic
expressions for small $N$ partition functions and Gaussian
approximations for them in the high temperature limit for all $N$. We
use numerical evidence to identify two temperature regimes where the
partition function has different limiting behaviors. The temperature
at which this transition takes place is identified as the Hagedorn
temperature.

\end{abstract}
\maketitle

\section{Introduction}

The string bit model was first introduced as a microscopic theory of
the lightcone quantized relativistic string \cite{giles_lattice_1977}.
In this model the string is seen as a polymer of more fundamental units
called string bits. To be more specific, each string bit carries an
infinitesimal unit of the $+$ component of momentum,
$P^+=\frac{P^0+P^1}{\sqrt{2}}$ of the string. As for $P^-$, it can be
identified as the Hamiltonian of a polymer of string bits. As the
number of bits, $M$, becomes large with a fixed $P^+$, the excitations
of this polymer of bits resemble those of a string with
$P^+ = m \mathcal{M}$, where $m$ is the unit of $P^+$ carried by each
bit and $\mathcal{M}$ is the bit number operator. These bits are
created from the vacuum state $|0\rangle$ by $N \times N$ matrix
creation operators, $\bar{\phi}_\rho^\sigma$, that impose a $U (N)$
color symmetry. Studying the large $N$ expansion
\cite{hooft_planar_1974} of the string bit dynamics, then, is
equivalent to doing string perturbation theory
\cite{thorn_reformulating_1994}. Although initially formulated to
describe bosonic strings, this formulation was soon extended to
superstrings \cite{bergman_string_1995}. The bit creation operator
was made completely antisymmetric in an additional set of $p$
``spinor'' indices $[u_1\cdots u_p]$ where $p\in\{0,s\}$ and
$u_q \in\{1,s\}$ with $s$ denoting the number of Grassmann worldsheet
fields in the emergent superstring. Depending upon the number of
spinor indices it has, a bit can be either bosonic or fermionic in
nature. The superstring bit model that we shall analyze is the simplest
possible one: where $s=1$. Recently, it was realized that the matrix
creation operator of superstring bits need not be a function of
transverse coordinates. Instead it is sufficient to devise another set
of two-valued internal ``flavor" degrees of freedom, $v_1\cdots v_d$,
where $v_q\in\{1,2\}$ with $d$ denoting the number of transverse
dimensions \cite{thorn_space_2014}. In this article we work with
space-less superstring bits: they have no space dependence. It is
important to realize that the string bits do not even have the
longitudinal dimension: that arises in the string theory limit. For
large $M$, $P^+$ becomes effectively a continuous momentum, giving
rise to its conjugate: the longitudinal coordinate.

The (canonical) partition function of a system at a thermal
equilibrium with a heat bath is defined as
\begin{dmath}
  Z (\beta) = \int^{\infty}_0 dE\ g(E) e^{- \beta E}
\end{dmath}
where $\beta$ is the reciprocal of the product of the Boltzmann
constant and the temperature and $g (E)$ is the density of microstates,
i.e. the number of states whose energies lie between $E$ and $E + d E$.
The partition function can also be regarded as the Laplace transform of
the density of states of the system. Also if the density of states
increases exponentially with $E$, the partition function can be seen to
diverge above a certain temperature (or, equivalently, below a certain
value of $\beta$). In statistical mechanics, partition functions can
only be singular in the thermodynamic limit. In first order phase
transitions, it is the first derivative of the free energy
($\propto \log(Z)$) that is discontinuous. This discontinuity that
occurs at the critical temperature, can be attributed to a latent heat
in the system. One way to think about a diverging partition function is
to imagine an infinite  latent heat of the system: it requires
adding infinite heat to change the temperature. Phase transitions
usually involve liberation of more degrees of freedom in the system
under consideration. Hagedorn was studying a thermodynamic model of
strongly interacting particles when he found out that for his model to
be consistent, the density of states of the system must grow
exponentially with energy \cite{hagedorn_statistical_1965}. This
suggested that there is a highest temperature that can be attained by
matter (or more specifically, hadrons). Later, dual resonance models
were shown to have such exponential dependence of the density of states
by Fubini and Veneziano \cite{fubini_level_1969}. This dual resonance
model has evolved into what is today known as string theory. In
\cite{atick_hagedorn_1988} the Hagedorn transition is seen as
liberating the underlying degrees of freedom in string theory.
 
This Hagedorn phenomenon was studied in the context of the superstring
bit model in \cite{thorn_string_2015}. The superstring bit model used
there has $s = 1$ and $d = 0$. This is the simplest superstring bit
model: there are only two kinds of creation operators:
$\left(\bar{\phi}\right)_\rho^\sigma$ and
$\left(\bar{\phi}_1\right)_\rho^\sigma$.
The former is bosonic (Grassmann-even) and the latter is fermionic
(Grassmann-odd) in nature. Since there are only two kinds of bits, we
can suppress the spinor indices on the operators, and instead  use the
symbol `a' for bosonic bits and `b' for fermionic bits. Thus, the only
indices that are left are the color indices. The Hamiltonian used in
\cite{thorn_string_2015} can be expressed as
\begin{dmath} H = \frac{T_0}{2 m N}  \text{tr} 
[(\bar{a}^2 - i\overline{b^{}}^2) a^2 - (\bar{b}^2 - i \bar{a}^2) b^2
 + (\bar{a} \bar{b} + \bar{b} \bar{a}) b a +
   (\bar{a} \bar{b} - \bar{b} \bar{a}) b a]
\label{eq:ham}
\end{dmath}
where, $T_0$ denotes the rest tension of the emergent string, `tr'
denotes the trace of the operators over color indices and other symbols
denoting the usual quantities we have already defined. The bosonic
annihilation operators are defined as 
$a_{\sigma}^{\rho} =(\bar{a}_{\rho}^{\sigma})^{\dagger}$ and these
have the following commutation relation:
\begin{dmath}
[ a_{\mu}^{\nu}, \bar{a}_{\sigma}^{\rho} ] = \delta_{\sigma}^{\nu}
   \delta^{\rho}_{\mu}
\end{dmath}
$\bar{b}_\rho^\sigma$ have similar definition for annihilation
operators, but they follow anti-commutation relation:
\begin{dmath}
\{b_{\mu}^{\nu}, \bar{b}_{\sigma}^{\rho}\} = \delta_{\sigma}^{\nu}
   \delta^{\rho}_{\mu}
\end{dmath}

In this paper we have investigated the onset of the Hagedorn phase
transition in our superstring bit model. One important feature of 
the energy spectrum of string bits under $H$ is that, in the large $N$
limit, the ground state energies of color singlet states and color
adjoint states are separated by a finite (and, in the limit
$M\rightarrow \infty$, constant) gap. However, the energy scale of
string excitations is of $O (1 / M)$ \cite{sun_stable_2014}.  One
way of interpreting this is that for the singlet states,
$H_{\text{eff}}$ is finite and, relatively speaking, for the adjoint
states, $H_{\text{eff}}=\infty$. In other words, this interaction
gives rise to color confinement: effectively, only the color singlet
states are significant in the appropriate limits. This means that
instead of using the full Hamiltonian and studying all the states of
the system, one may study the Hagedorn phenomenon in a system in which
the only dynamics is singlet restriction. This is like imposing color
confinement by hand, instead of letting it emerge on its own. As we
shall see, this drastically simplified system is still
rich enough to support a Hagedorn phenomenon. 

The thermal perturbation scheme developed in \cite{thorn_string_2015}
is well defined for arbitrary large values of temperature, and yet 
the density of singlet states at large $N$ \cite{chen_numerical_2016}
suggest a finite limiting temperature. This suggests that the Hagedorn
phenomenon is an artifact of the large $N$ limit. In this paper we
study the superstring bit model at finite $N$ with $H=0$ and
singlet restrictions in order to understand the source of the Hagedorn
phenomenon at $N\rightarrow \infty$. We present some exact results for
low $N$ partition functions and approximations in the high temperature
limit. We also obtain numerical data and analyze them to identify and
establish the difference in the behavior of superstring bits below and
above the Hagedorn temperature.

\section{Counting of singlets at large N}
Let us set up the problem that we are studying in this paper. We
have already described the model that we are studying. We have also
explained that we are using $H=0$ and restricting ourselves to
the singlet sector. As we are studying the thermal properties of the
system, we shall be working with the canonical partition function
for a system of superstring bits. In terms of light-cone parameters,
\begin{align}
P^- &=H=0\nonumber\\
\implies P^0&=\frac{P^+ +P^-}{\sqrt{2}}\nonumber\\
&=\underbrace{\frac{m}{\sqrt{2}}}_{=\ \omega} \mathcal{M}
=\omega\ \text{tr}(\bar{a} a +\bar{b} b)
\label{eq:omega}
\end{align}
Our partition function would then be defined as
$Z=\text{Tr}\exp(-\beta P^0)$ where `Tr' denotes the thermal trace,
i.e. the trace over all the singlet eigenstates.
Since, $H\propto T_0$ (Eq.~\ref{eq:ham}), the $H=0$ limit can also be
regarded as the tensionless limit of the emergent string.

\subsection{Only bosonic bits}
Let us derive the large $N$ partition function when there is only
one bosonic oscillator. For this, a basis element for color singlets
can be written, up to a normalization constant, as:
\[ |  \text{singlet} \rangle_{ \text{bos}} =   \text{tr} (\bar{a})^{l_1}   \text{tr}
   (\bar{a}^2)^{l_2} \cdots | 0 \rangle =\left( \prod_{r = 1}^{\infty}   \text{tr}
   (\bar{a}^r)^{l_r}\right)  | 0 \rangle \]
where each $l$ can be any non-negative integer. In fact, there is a
one-to-one correspondence between the set of all such singlets and the
set of sequences ($l_1,l_2,\cdots$). Then, energy of such a state is
 given by:
\begin{align}
P^0 |  \{l_r\} \rangle_{ \text{bos}} &= \omega \left( \sum_{r =   1}^{\infty} r l_r \right)  | \{l_r\} \rangle_{ \text{bos}}\nonumber\\
   \implies e^{- \beta P^0} | \{l_r\} \rangle_{ \text{bos}} &= \left(   \prod_{r = 1}^{\infty} e^{- \beta \omega r l_r} \right)  | \{l_r\} \rangle_{ \text{bos}}
\end{align}
where $\omega$ is defined as in Eq.~\ref{eq:omega}. In the large $N$
limit all these states form an orthogonal eigenbasis of the system.
Hence, in that limit, the singlet partition function for pure bosons
is:
\begin{dmath}
\label{z_in_bos}
\lim_{N\rightarrow \infty}Z_{ \text{bos}}(\beta) 
= \sum_{\{ l_1, l_2, \cdots \}} \prod_{r =  1}^{\infty} e^{- \beta \omega r l_r}
= \prod_{r = 1}^{\infty} \frac{1}{1 -   e^{- \beta \omega r}}
\end{dmath}
\subsection{Only fermionic bits}
For one fermionic oscillator, one has to be careful, because some of
the single trace operators are simply zero. This is  because of the
cyclic property of the matrix trace and the anti-commutativity of the
$\bar{b}$s. E.g.
\[ \text{tr} (\bar{b}^2) =  -  \text{tr} (\bar{b}^2) = 0\]
In fact, all trace operators with an even number of bits
are zero. Hence, a basis element for fermion singlets looks like
\begin{dmath*}
 |   \text{singlet} \rangle_{ \text{fer}} =   \text{tr} (\bar{b})^{k_1}   \text{tr}
   (\bar{b}^3)^{k_3} \cdots | 0 \rangle = \left(\prod_{r = 1}^{\infty}   \text{tr}
   (\bar{b}^{2 r - 1})^{k_{2 r - 1}}\right) | 0 \rangle
\end{dmath*}
where, each $k$ can take only two values: $0$ and $1$. Hence, for the
large-$N$ partition function, we have:
\begin{dmath}
\label{z_in_fer}
\lim_{N\rightarrow \infty} Z_{ \text{fer}}(\beta) = \sum_{\{ k_1, k_3, \cdots \}=0}^1 \ \prod_{r =   1}^{\infty} e^{- \beta \omega (2 r - 1) k_{2 r - 1}}
= \prod_{r =   1}^{\infty} \{1 + e^{- \beta \omega (2 r - 1)}\}
\end{dmath}

\subsection{Supersymmetric case}
\label{sec:SupersymmetricCase}
One may verify that neither $Z_{\text{bos}}$ nor $Z_{\text{fer}}$
diverges at a finite temperature. However, in the supersymmetric
case, it can be shown that there is an exponential degeneracy in the
number of singlets \cite{chen_numerical_2016}, and thereby a finite
Hagedorn temperature.
Naively, this can be understood as follows: each bit can be either
bosonic or fermionic, hence there are roughly $2^M/M$ possibilities for
a single trace operator of $M$ supersymmetric bits (multi trace states
are singlets as well). Of course, some of these combinations don't
correspond to physical states because of the arrangement of
anti-commuting fermionic operators in them. E.g.
\[
\text{tr}(\bar{b}\bar{a}^2\bar{b}\bar{a}^2)
=\text{tr}(\bar{b}\bar{b}\bar{a}^2\bar{a}^2)
=-\text{tr}(\bar{b}\bar{a}^2\bar{b}\bar{a}^2)=0
\]
However, in the large $M$ limit, this doesn't harm the
exponential degeneracy of eigenstates. In fact, the degeneracy of
single trace states goes as $ 2^{M-1}/M$ for large $M$ \cite{chen_numerical_2016,aharony_hagedorn/deconfinement_2004,sundborg_hagedorn_2000}.
One may calculate an approximation to the partition function starting
from this degeneracy. It turns out that the exact supersymmetric
partition function,
\begin{equation}
\lim_{N\rightarrow \infty}Z(\beta)=\prod_{n=1}^\infty {\frac{1}{1-2\ \exp\{-\beta \omega (2n-1)\}}}
\label{eq:z_susy}
\end{equation}
A complete derivation of the generalized result and some
interesting sub-cases will appear in \cite{raha_thorn}. 
As is evident from Eq.~\ref{eq:z_susy}, $Z$ diverges at
$\exp\{-\beta \omega\}=1/2$ (it also diverges at other, larger values
of temperature). Hence, as anticipated, the supersymmetric case has a
Hagedorn temperature at $T_H=\omega/\log(2)$. We shall come back to
this result in section \ref{sec:int}.

All the results quoted in this section are applicable in the large $N$
limit. In this paper we wish to study finite $N$ partition functions.
Hence, we have to develop a systematic method of counting the
eigenstates (i.e. singlets that are linearly independent) for this
model at finite $N$. This is what we shall do in the next section.

\section{From characters to partition functions}

Imposing the singlet restriction on the Fock space of all physical
states is a purely group theoretic exercise. In our case, the group
under consideration is $SU(N)$, the gauge group of the model. A bit
creation operator has two color indices and transforms under the
adjoint action of $SU(N)$. Given a number of such adjoint operators
we shall count the number of ways in which one may obtain states
that transform trivially under $SU(N)$. This is very similar to
decomposing a direct product of representations into a direct sum of
irreps.

Before proceeding with the supersymmetric case, let us examine the pure
bosonic and pure fermionic cases at finite $N$. The partition functions
are
\begin{align}
Z_{\text{bos}}(N,\beta)&=\prod_{r = 1}^{N} \frac{1}{1 -   e^{-\beta \omega r}}\label{z_fin_bos}\\
Z_{\text{fer}}(N,\beta)&=\prod_{r = 1}^{N} \{1 +   e^{-\beta \omega (2r-1)}\}\label{z_fin_fer}
\end{align}
As one can see, these expressions are similar to $N\rightarrow \infty$
ones, except that the product over $r$ has been truncated at $r=N$.
This is because when $N$ is finite, for $L> N$, $\bar{a}^L$ is linearly
dependent on $\bar{a}^c$s for $c\in\{1,2,\cdots, N\}$. One can obtain
the exact dependence from the Cayley-Hamilton theorem, which also
gives trace identities for different powers of any square matrix
over a commutative ring. For the fermionic case, one has to use a
generalized version of the Cayley-Hamilton theorem in order to obtain
the cutoff. However, there is another explanation for the fermionic
result: it is the Poincar\'e polynomial for $SU(N)$
(see, for example, ch \RN{7} sec 11 of \cite{weyl}). The coefficients
of powers of $e^{-\beta \omega}$ in this polynomial count the number of
invariants that are linear but completely antisymmetric in the
infinitesimal elements of $SU(N)$.

Now that we have stated the special cases, let us focus on the task at
hand. For this purpose, we shall use group characters. We shall make
use of the orthogonality of characters of different irreps of a group.
More specifically, the idea is to obtain the representational content
of a physical state by writing down its character. Given this
character, one may extract the multiplicity of an irrep  within it by
taking its product with the conjugate of the character of the said
irrep, integrating this over the entire group and then normalizing
the integral:
\begin{dmath}
  g_{RR'} = \frac{\int_G \{\prod_i d \theta_i\}\ h (\{ \theta_i \})\ \chi^{\ast}_R (\{  \theta_i \})\  \chi_{R'} (\{ \theta_i \})}{\int_G \{\prod_i d \theta_i\}\ h (\{  \theta_i \}) }
\label{eq:char}
\end{dmath}
where $g_{RR'}$ denotes the multiplicity of the irrep $R$ in the
reducible representation $R'$, $h (\{ \theta_i \})$ denotes the Haar
measure on the group G and $\chi_{R} (\{ \theta_i \})$ represents
the character of the representation $R$. All these quantities are
parametrized in terms of \{$\theta_i$\}; the rotation angles
corresponding to the Cartan subalgebra of the group G. This is very
much like using a
projection operator to extract out the subspace of irrep $R$.
Instead of the character for only one state, one can use a
suitable character generating function and obtain the corresponding
multiplicity generating function. In \cite{curtright_symmetry_1986}
such a generating function was derived to obtain multiplicities at
arbitrary mass levels of strings. 

For $SU(N)$, the Haar measure is given by
\begin{dmath} \prod_{1 \leqslant i < j \leqslant N} | e^{i \theta_j} - e^{i \theta_i} |^2 =
   \prod_{1 \leqslant i < j \leqslant N} 4 \sin^2 \left( \frac{\theta_i - \theta_j}{2}
   \right) \end{dmath}
Also, the character for a singlet, the irrep we seek, is simply $1$.
Let us figure out how to write down the character for a given state.
As mentioned earlier, a creation operator, $\bar{a}^\mu_\nu$ or $\bar{b}^\mu_\nu$,
transforms in the adjoint representation of $SU(N)$. As such, its
upper and lower indices can be regarded as transforming in the
fundamental and the anti-fundamental representations, respectively.
The character of an operator transforming in the fundamental
(anti-fundamental) representation is just $\exp\{i\theta_k\}$ 
($\exp\{-i\theta_k\}$) where $\theta_k$ is the rotation angle
corresponding to it. Hence, the character for a state
$\bar{a}_\nu^\mu|0\rangle$ or $\bar{b}_\nu^\mu|0\rangle$ is given by
$\exp\{i(\theta_\mu-\theta_\nu)\}$. In doing this exercise, we have
made an oversimplification: the adjoint irrep isn't strictly the
direct product of the fundamental and the anti-fundamental irreps. We
shall compensate for this error in a moment.

Given a pair of color indices, ($i,j$), an arbitrary multiplet can
be represented as
\[\left(\bar{a}^i_j\right)^{l_{ij}}\left(\bar{b}^i_j\right)^{k_{ij}}|0\rangle\]
where $l_{ij}$ can be any non-negative integer and $k_{ij}$ can be
either $0$ or $1$. From this, one may construct a character generating
function for the all states with colors ($i,j$):
\begin{dmath}
\frac{1+x e^{i(\theta_i-\theta_j)}}{1-x e^{i(\theta_i-\theta_j)}}=\left(1+x e^{i(\theta_i-\theta_j)}\right)\left(1+x e^{i(\theta_i-\theta_j)}+x^2 e^{2i(\theta_i-\theta_j)}+\cdots \right)
\end{dmath}
The numerator above counts the contribution from the
fermionic bits while the denominator refers to the contributions
coming from the bosonic bits. Hence, the character generating
function of any multiplet of any number of superstring bits is given
by the expression
\begin{dmath}\prod_{1\leqslant i,j\leqslant N}
    \frac{1 + x e^{i (\theta_i - \theta_j)}}{1 - x e^{i (\theta_i -
   \theta_j)}} = \left( \frac{1 + x}{1 - x} \right)^{N} \prod_{1
   \leqslant i < j \leqslant N} \frac{1 + x^2 + 2 x \cos (\theta_i - \theta_j)}{1 + x^2
   - 2 x \cos (\theta_i - \theta_j)}
\label{eq:un}
\end{dmath}
So far in this derivation we have approximated the adjoint irrep as a
direct product of fundamental and anti-fundamental irreps. However, 
this product includes an additional irrep: the $U(1)$ singlet.
Eq.~\ref{eq:un}, technically speaking, gives the $U(N)$ character of
our model. In order to obtain the generating function for $SU(N)$ we
have to remove the tr$(\bar{a})$ from each $\bar{a}_\nu^\mu$ and
tr$(\bar{b})$ from each $\bar{b}_\nu^\mu$.
\[
\left( \frac{1 + x}{1 - x} \right)^{N-1} \prod_{1
   \leqslant i < j \leqslant N} \frac{1 + x^2 + 2 x \cos (\theta_i - \theta_j)}{1 + x^2
   - 2 x \cos (\theta_i - \theta_j)}
\]
The coefficient of $x^M$ in the expression above gives the correct
$SU(N)$ character of the subspace of states with bit number $M$. If we
cast this character generating function into Eq.~\ref{eq:char} we
obtain the generating function for multiplicities of singlet states
\begin{widetext}
\begin{dmath}
  \sum_{M=0}^\infty{{g_M} \  x^M} = \left( \frac{1 + x}{1 - x} \right)^{N - 1}
  \frac{\int^ \pi_{-\pi}\left( \prod^N_{k = 1} d \theta_k\right) \prod_{1 \leqslant i < j
  \leqslant N}  4 \sin^2 \left( \frac{\theta_i - \theta_j}{2}\right)  \frac{1 + x^2 + 2 x \cos
  (\theta_i - \theta_j)}{1 + x^2 - 2 x \cos (\theta_i - \theta_j)}}{\int_{-\pi}^\pi \left(\prod^N_{k = 1} d \theta_k \right) \prod_{1 \leqslant i < j \leqslant N} 4 \sin^2 \left( \frac{\theta_i - \theta_j}{2}\right)}
\label{eq:original}
\end{dmath}
\end{widetext}

The R.H.S. of this equation has the following properties:-
\begin{enumerate}
\item The integrand in the denominator is the $x \rightarrow 0$ (low
temperature) limit of the integrand in the numerator. It can also be
interpreted as the ``volume'' of the $SU(N)$ group and evaluates to
$N! (2 \pi)^N$.
\item The integrand is completely symmetric in the \{$\theta_k$\}s.
\item It is also periodic in each $\theta_k$. Hence, the domain of
integration is $\mathbb{T}^N$, i.e. the $N$-torus.
\item The integrand is a function of differences of $\theta_k$'s; in
fact it includes all the $^N C_2$ differences among the $\theta_k$
's. This tells us that the integral is translation-invariant in the
$\theta_k$'s.
\item The integrand (or rather, the Haar measure) vanishes if
$\theta_i=\theta_j$ for any $i\neq j$.
\end{enumerate}

This generating function gives the multiplicities of singlets at
any bit number. In our model, $E=\omega M$ since, $H=0$. If we
identify $x$ in this function with $\exp\{-\beta \omega\}$, the
L.H.S. of Eq.~\ref{eq:original} becomes the (canonical) partition
function.
\[
\sum_{M=0}^\infty{g_M\ \exp\{-\beta \omega M\}}=Z(N,\beta)
\]
Thus, all partition functions mentioned so far (Eqs.~\ref{z_in_bos},~\ref{z_in_fer},~\ref{eq:z_susy},~\ref{z_fin_bos},~\ref{z_fin_fer}),
are just different generating functions of $U(N)$ characters. In order
to obtain the corresponding $SU(N)$ characters, one must multiply them
with $\frac{1-x}{1+x}$. From here onwards, whenever we mention
partition functions we shall be referring to generating functions of
$SU(N)$ characters. Also for simplicity, we shall express $Z$ as a
function of $x(=\beta \omega)$ as opposed to $\beta$. 

\section{High Temperature Limit of Z}
\begin{figure}[htb]
	\centering
		\includegraphics[width=0.80\textwidth]{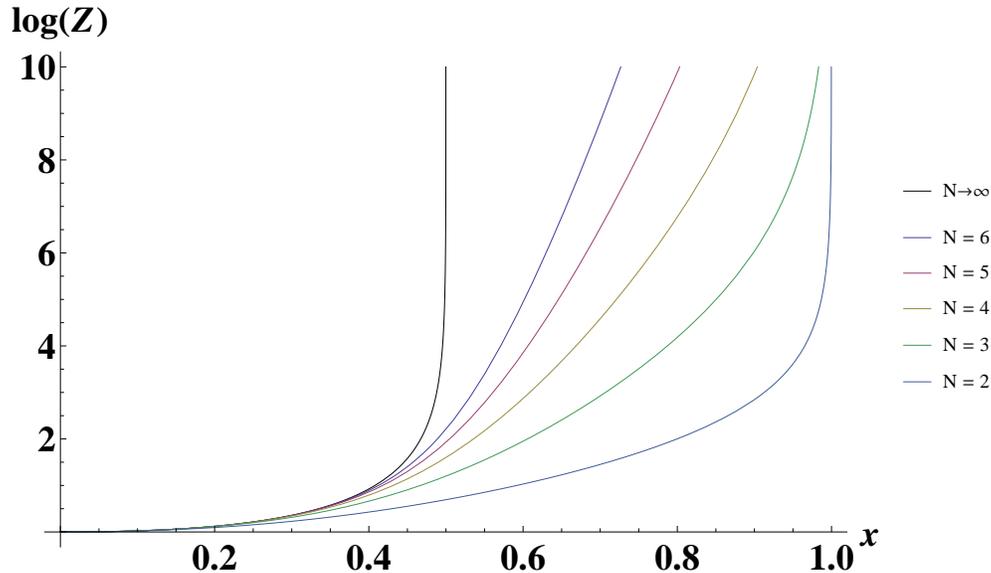}
	\caption{Temperature dependence of known partition functions}
	\label{fig:exact_z}
\end{figure}

The partition functions for some values of $N$ have been tabulated
in TAB.~\ref{tab:Exact} and plotted in FIG. \ref{fig:exact_z}. Beyond
$N = 2$, the character integral becomes too involved to be done
analytically by hand. We used a computer to obtain results for $N>2$.
Beyond $N=3$, it is useful to employ a change of variables
$e^{i\theta_j} \rightarrow z_j$, in order to turn this integration
problem into a problem of calculating (multidimensional) residues:
\begin{widetext}
\begin{dmath} Z(N,x) = \frac{1}{N!}  \left( \frac{1 + x}{1 - x}
   \right)^{N - 1} \sum_{\mathbb{T}^N}  \text{Res} \left( \prod_{1 \leqslant i < j
   \leqslant N} \frac{- (z_i - z_j)^2 (x z_i + z_j) (x z_j + z_i)}{ (x z_i -
   z_j) (x z_j - z_i)}  \prod_{1 \leqslant k \leqslant N} \frac{1}{(z_k)^N}
   \right)
\end{dmath}
\end{widetext}
However, even calculating these residues is prohibitively time-consuming for $N > 6$.
\begin{table*}[htb]
	\centering
		\begin{ruledtabular}    
           \begin{tabular}{|C{1cm}|C{15cm}|}
            $N$&$Z(x)$\\
            \hline
            2 &$(1 - x + 2 x^2)/(1 - x)$\\
            \hline
            3 &$(1 + x) (1 - 2x + 4 x^2 - 3 x^3 + 4 x^4 + 2 x^5 + 4 x^7)/\left[(1 - x)^2  (1 + x + x^2)\right]$\\
            \hline
            4 & $(1 + x) (1 - 3 x + 6 x^2 - 7 x^3 + 9 x^4 - 6 x^5 + 10 x^6 - 2 x^7 +  12 x^8 + 8 x^9 + 8 x^{10} + 16 x^{11} + 8 x^{12} + 8 x^{13} + 8 x^{14})/\left[(1   - x)^3  (1 + x + x^2)\right]$ \\
            \hline
            5 & $(1 + x)^2 (1 - 4 x + 10 x^2 - 17 x^3 + 26 x^4 - 31 x^5 + 40 x^6 - 36 x^7 + 49 x^8 - 18 x^9 + 42 x^{10} + 52 x^{11} + 38 x^{12} +  148 x^{13} + 108 x^{14} + 240 x^{15} + 244 x^{16} + 344 x^{17} + 376 x^{18} +  392 x^{19} + 448 x^{20} + 352 x^{21} + 360 x^{22} + 272 x^{23} + 176 x^{24} + 144 x^{25} + 64 x^{26} + 32 x^{27} + 16 x^{28})/\left[(1- x)^4 (1 + x + x^2) (1 + x + x^2 + x^3 + x^4)\right]$\\
            \hline
            6 & $(1 + x)^2 (1 - 4 x + 10 x^2 - 18 x^3 + 30 x^4 - 41 x^5 + 59 x^6 - 68 x^7 + 98 x^8 - 84 x^9 + 140 x^{10} - 41 x^{11} + 206 x^{12} + 128 x^{13} + 442 x^{14} + 572 x^{15} + 1130 x^{16} + 1764 x^{17} + 2824 x^{18} + 4468 x^{19} + 6616 x^{20} + 9712 x^{21} + 13688 x^{22} + 18656 x^{23} + 24488 x^{24} + 31152 x^{25} + 38016 x^{26} + 44632 x^{27} + 50640 x^{28} + 54792 x^{29} + 57120 x^{30} + 57056 x^{31} + 54368 x^{32} + 49632 x^{33} + 43232 x^{34} +35776 x^{35} + 28160 x^{36} + 21088 x^{37} + 14816 x^{38} + 9824 x^{39} + 6144 x^{40} + 3488 x^{41} + 1856 x^{42} + 896 x^{43} + 352 x^{44} + 128 x^{45} + 32 x^{46})/\left[(1-x)^5 (1+x+x^2)^2 (1+x+x^2+x^3+x^4)\right]$\\
            \hline
            $\vdots$ & $\vdots$\\
            \hline
            $\infty$ &$ (1-x)/\left[(1+x)\ \prod_{k=1}^\infty( 1-2x^{2k-1})\right]$
           \end{tabular}
        \end{ruledtabular}
	\caption{Partition functions that are known exactly.}
	\label{tab:Exact}
\end{table*}

We have already identified the $x \rightarrow 0$ limit of
Eq.~\ref{eq:original} : it is when the R.H.S. becomes $1$ as the
numerator becomes equal to the denominator. Let us now concentrate
on the $x \rightarrow 1$ (high temperature) limit.
\begin{dmath}
Z(N,x\rightarrow 1)=\left( \frac{2}{1 - x}
  \right)^{N - 1} \frac{\int^{\pi}_{-\pi} \prod^N_{k = 1} d \theta_k
 \prod_{1  \leqslant i < j \leqslant N} 4 \cos^2 
\left(\frac{\theta_i - \theta_j}{2}\right) }{N! (2  \pi)^N}
\label{eq:high}
\end{dmath}
As before, evaluating this integral analytically isn't straightforward
beyond the first few values of $N$. The results for $N = 2-5$ can be
trivially obtained. Beyond $N=5$, one has to use the corresponding
multidimensional residue form 
\[\left( \frac{2}{1 - x} \right)^{N - 1} \frac{R_N}{N!}\]
where
\begin{dmath}
 R_N=\text{Res}  \left(  \frac{\prod_{1 \leqslant i < j \leqslant N}  (z_i +  z_j)^2}{\prod_{1 \leqslant k \leqslant N}  (z_k)^N} \right)
\label{eq:rn}
\end{dmath}
This has only one pole (of order $N$) enclosed in the domain of
integration; it is at the origin, $(0,\cdots,0)$. For evaluating $R_N$,
it can also be interpreted as the coefficient of the term
$\prod_{1 \leqslant k\leqslant N}\  (z_k)^{N - 1}$
in the Taylor-series expansion of
$\prod_{1\leqslant i < j \leqslant N}\  (z_i + z_j)^2$
about the origin of the $\mathbb{T}^N$. This term has every $z_k$
raised to the same power ($N-1$), consequently, its coefficient is
the biggest of all the coefficients in the Taylor series. Here too,
beyond $N>7$ the computation becomes increasingly time-consuming.
A list of the calculated $R_N$ values can be found in
TAB.~\ref{tab:rn}.
\begin{table}[htb]
	\centering
        %\begin{ruledtabular}
        	\begin{tabular}{|C{1cm}|C{3cm}|}
                \hline
                N&$R_N$\\
                \hline
                2&2\\
                3&10\\
                4&152\\
                5&7736\\
                6&1375952\\
                7&877901648\\
                \hline
            \end{tabular}
        %\end{ruledtabular}
	\caption{Exact values of $R_N$.}
	\label{tab:rn}
\end{table}

\subsection*{Asymptotic Behavior: Steepest Descent}
Since, we don't have a closed form expression for different
coefficients of $\prod_{1\leqslant i < j \leqslant N}\  (z_i + z_j)^2$,
we turned to its asymptotic analysis. Examining the expression $R_N$
we see that it is clearly greater than $2^{N (N - 1) / 2}$, as the
latter is the coefficient of the term
$\prod_{1\leqslant k \leqslant N} (z_k)^{N - 1}$
in the Taylor series of
$\prod_{1\leqslant i < j \leqslant N} 2 z_i z_j$.
Similarly, $4^{N (N_{} - 1) / 2} = 2^{N (N - 1)}$ is a conservative
upper bound on $R_N$, as it is the sum of all coefficients in
the expansion of the
$\prod_{1 \leqslant i < j \leqslant N} (z_i + z_j)^2$.
With these two bounds we can deduce the $N$ dependence of $R_N$: it
goes as $\exp(N^2)$ in the leading order. This is an important
feature and we shall refer to this observation later.
Examining Eq.~\ref{eq:high}, we see that the integrand attains its
maximum value when all the $\theta$s are equal to each other. If one
fixes the value of, say $\theta_1$, to be $\psi$; the modified
integrand attains a single, isolated global maximum at
$\theta_2=\theta_3=\cdots=\theta_N=\psi$. While this modification
doesn't change the value integral (or $R_N$), but it allows for a
simpler, more accurate way of approximating it: by the method of
steepest descent.

The method of steepest descent is well known for approximating
integrations with a fixed number of dimensions. In our case, however,
the number of dimensions isn't fixed, it's increasing. In
fact, $N$ itself is the large parameter in our case. This isn't
obvious as integrand has an implicit dependence on $N$.

Let us express the general integrand in Eq.~\ref{eq:original} as
$\exp\left\{\widehat{f}(x,\{\theta_k\})\right\}$, where
\begin{dmath}
\widehat{f}(x,\{\theta_k\}) = \sum_{1\leqslant i<j\leqslant N}\log \left\{ 4 \sin^2 \left(\frac{\theta_i - \theta_j}{2}\right)  \frac{1 + x^2 + 2 x \cos(\theta_i - \theta_j)}{1 + x^2 - 2 x \cos (\theta_i - \theta_j)} \right\}
\label{eq:gauss}
\end{dmath}
For large $N$, one can convert this restricted double sum into the
Cauchy principle value of a double integral. We introduce an
integration variable $y_i$ such that
$y_i=\frac{i}{N}\implies dy_i=\frac{\Delta i}{N}=\frac{1}{N}$
and a non-decreasing function $\Theta(y)$ for $y\in (0,1]$ such that
$\Theta(y_i)=\theta_i$.
While defining $\Theta$ we have taken advantage of the fact that in
$\widehat{f}(x,\{\theta_k\})$, the order of the sequence of $\theta$s
doesn't matter. Hence, one can always redefine the $\theta$s such
that they form a non-decreasing sequence, thereby allowing us to
define a unique function in the continuum limit. Now,
$\widehat{f}(x,\{\theta_k\})$ can re-written as
\begin{widetext}
\begin{dmath}
\widehat{f}(x,\{\theta_k\}) = \widehat{f}(N,x)
= N^2 \lim_{\epsilon\rightarrow 0}\int_0^1 \int_0^{y-\epsilon}\log
\left\{ 4 \sin^2 \left(\frac{\Theta(y') - \Theta(y)}{2}\right) 
\frac{1 + x^2 + 2 x \cos(\Theta(y') - \Theta(y))}{1 + x^2 - 2 x
\cos (\Theta(y') - \Theta(y))} \right\}dy'dy
\label{eq:N2}
\end{dmath}
\end{widetext}

\begin{figure}[htb]
	\centering
		\includegraphics[width=0.80\textwidth]{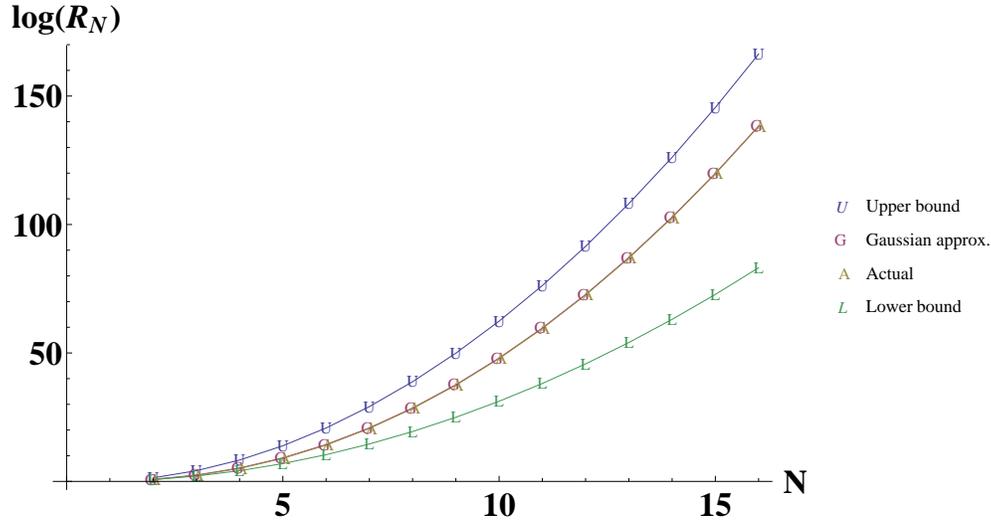}
	\caption{Different estimates of $R_N$ at infinite temperature. The exact values for $N>7$ were obtained from \cite{mckay_applications_1983}}
	\label{fig:hightemp}
\end{figure}

As can be seen above, going to the continuum limit brings out an
overall factor of $N^2$, thereby making the dependence of
$\widehat{f}(x,\{\theta_k\})$ on $N$ explicit. Now that we have
justified applying the method of steepest descent to our problem, we
can see what it yields in the high temperature limit. It gives 
\begin{dmath}
R_N=\frac{2^{N (N -1)}}{\pi^{(N - 1) / 2} N^{N / 2 - 1}}
\label{eq:x1val}
\end{dmath}
which is a more stringent upper bound than $2^{N(N-1)}$. A detailed
derivation of this result may be found in the appendix. All these
approximations have been plotted against known exact values in
FIG.~\ref{fig:hightemp}. 

Another useful way of expressing $\widehat{f}(N,x)$ is:
\begin{widetext}
\begin{dmath}
\widehat{f}(N,x)= N^2 \lim_{\delta\rightarrow 0}\int_0^{2\pi} \int_0^{\Theta-\delta}\log \left\{ 4 \sin^2 \left(\frac{\Theta' - \Theta}{2}\right)  \frac{1 + x^2 + 2 x \cos(\Theta' - \Theta)}{1 + x^2 - 2 x \cos (\Theta' - \Theta)} \right\}\rho(\Theta')\ d\Theta'\rho(\Theta)\ d\Theta
\label{eq:dist}
\end{dmath}
\end{widetext}

where we have introduced a normalized density function, $\rho(\Theta)$
such that $dy=\rho(\Theta)\ d\Theta$ and $\int_0^{2\pi}{\rho(\Theta)\ d\Theta}=1$.
Now, the global maxima of $\widehat{f}(x,\{\theta_k\})$ contribute most
to the Gaussian approximation. Hence, the density function of the
global maxima are of particular interest while doing the integral. The
stationarity condition for $\widehat{f}(x,\{\theta_k\})$, for $k\in\{1,2,\cdots,N\}$:
\begin{dmath}
\sum_{l\neq k}{\cot\left(\frac{\theta_k-\theta_l}{2}\right)-\frac{4x(1+x^2)\sin(\theta_k-\theta_l)}{1+x^4-2x^2\cos\{2(\theta_k-\theta_l)\}}}=0
\end{dmath}
If $\{\psi_k\}$ is a global maximum, then the density function,
$\rho(\Psi)$ must be a  solution of:
\begin{dmath}
\fint_{-\Psi_{[\text{max}]}}^{\Psi_{[\text{max}]}}{\left\{\cot\left(\frac{\Psi'-\Psi}{2}\right)-\frac{4x(1+x^2)\sin(\Psi'-\Psi)}{1+x^4-2x^2\cos\{2(\Psi'-\Psi)\}}\right\}\rho(\Psi)\ d\Psi}=0
\label{eq:maxro}
\end{dmath}
where $\fint$ denotes the principal value of the integral and the
limits are such that $\rho\left(\pm\Psi_{[\text{max}]}\right)=0$ and
$\Psi_{[\text{max}]}\leqslant \pi$. This technique follows from
\cite{brezin_planar_1978} where the authors study the Hermitian matrix
model. They used this technique to obtain an analytic expression for
the density function of the eigenvalues in the presence of quartic and
cubic interactions. We shall come back to these density functions later in the article.

\section{Intermediate Temperatures at Finite N}
\label{sec:int}
As we have already mentioned, the large $N$ energy spectrum for
superstring singlet states predicts a divergence in $Z$ at $x = 1/2$.
In order to trace the roots of this phenomena to finite $N$ partition
functions, one has to compare the temperature dependence (or, $x$
dependence) of $Z(N,x)$ below and above $x = 1/2$. At finite $N$,
the partition function is smooth over the entire temperature
range (FIG.~\ref{fig:exact_z}). This observation is confirmed by the
exact analytic expressions obtained for the first few $N$ values
(TAB.~\ref{tab:Exact}). Instead of studying temperature dependence of
the partition functions, one can fix $x$ (or, equivalently,
temperature) and analyze the values of $Z(N,x)$ as a sequence in $N$.
If the method of characters is correct then this sequence must
culminate in the appropriate $N\rightarrow \infty$ value (which is
known). In other
words, at any temperature below $x = 1/2$ this sequence must approach
the value $\lim_{N \rightarrow \infty}Z(x)$. And above $x =1/2$ this
sequence must diverge. This is the aim for our study: to establish
the existence of a low-temperature regime, where $\log (Z)$ has a
limit point and a high-temperature regime, where it does not have a
limit point.

Having defined our goal, we set out to evaluate $\log (Z)$ analytically
for finite values of $N$. However, as already mentioned before, that
exercise is time-consuming on our computer for $N > 5$. Thereafter,
encouraged by the success of the steepest descent method, we attempted
to extend it beyond $x= 1$. However, at general $x$, there are multiple
global maxima and the Hessian matrix develops a complicated
dependence on $x$. Both these factors prevent an analytic Gaussian
approximation of the integral at hand. Numerically, Monte-Carlo
method is the only hope for achieving an acceptable degree of
precision and accuracy for this multidimensional integral. However,
beyond $N = 7$, even Monte-Carlo method is no longer robust.
Numerical methods aren't feasible for evaluating the Gaussian
approximations either: beyond the first few values of $N$, calculating
the Hessian matrix isn't simple even numerically.

Thereafter, we chose to concentrate on the global maxima, and extract
as much information as we could, rather than continue pushing for the
higher order fluctuations. As we shall show in the following plots and
paragraphs, we were able to obtain evidence of the contrasting
behavior of the partition function at different temperatures.

Here is a description of our numerical study:
\begin{enumerate}
	\item We defined the following function:
    \begin{dmath}\widetilde{f}(N,x,\{\theta_k\})=\widehat{f}(x,\{\theta_k\})
    +(N-1)\log\left\{\frac{1+x}{1-x}\right\}
    \end{dmath}
    We performed a global maximization routine on $\widetilde{f}(N,x,\{\theta_k\})$.
    Since $\widetilde{f}(N,x,\{\theta_k\})$ is
    translation-invariant in
    $\theta$s, we were able to simplify this exercise slightly by
    fixing $\theta_1$ to $0$. This reduced the search space from
    $(-\pi,\pi)^N$ to $(-\pi,\pi)^{N-1}$. This maximization was
    repeated for a set of ($N,x$) pairs, namely, for $N\in\{9,13,\cdots,101\}$ and $x\in\{0, 1/40,\cdots, 39/40\}$.
    
    \item We obtained the coordinates, $\{\widetilde{\psi}_k\}_{N,x}$
    and the value, $\widetilde{f}^{[\text{max}]}_{N,x}$ of the global
    maximum for each $(N,x)$. The $\{\widetilde{\psi}_k\}_{N,x}$ 
    were sorted; and thereafter, shifted, to obtain a non-decreasing 
    sequence $\{\psi_k\}_{N,x}$ that is centered at $0$. The maximum 
    values were redefined to obtain
    \[f^{[\text{max}]}_{N,x}=\widetilde{f}^{[\text{max}]}_{N,x}-\widetilde{f}^{[\text{max}]}_{N,0}\]
    These $f^{[\text{max}]}_{N,x}$ resemble the function $\log{Z(N,x)}$
    more closely, e.g. $f^{[\text{max}]}_{N,0}=0=\log(Z(N,0))$ for all
    values of $N$. We discuss the corrections to this approximation in
    the appendix \ref{sec:deriv}. FIG.~\ref{fig:max_vs_x} shows 
    temperature dependence of $f^{[\text{max}]}$ for some values of
    $N$.    
\begin{figure}[htb]
	\centering
		\includegraphics[width=0.80\textwidth]{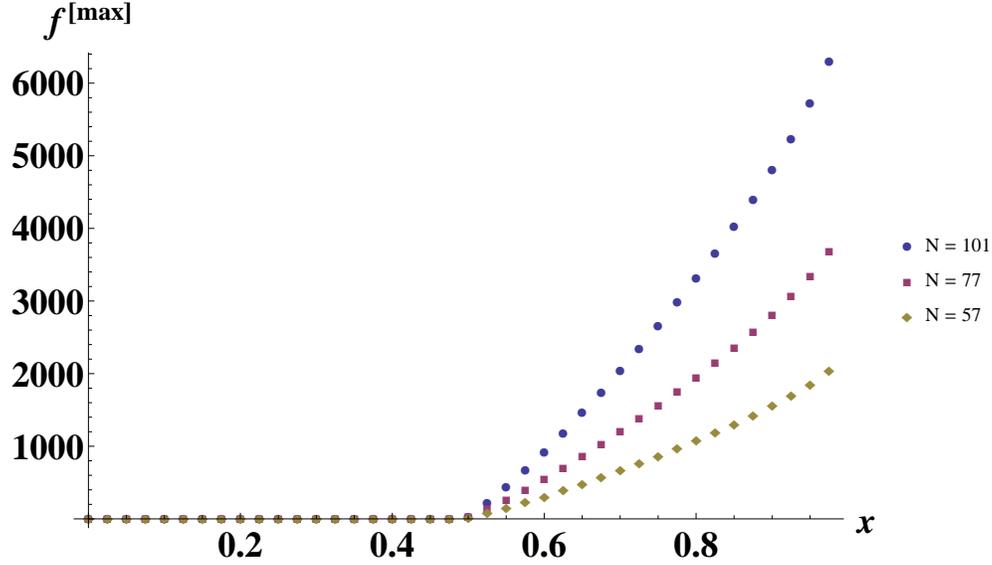}
	\caption{Dependence of $f^{[\text{max}]}$ on temperature.}
	\label{fig:max_vs_x}
\end{figure}
    
    \item We used $f^{[\text{max}]}_{N,x}$ to extract the dependence
    of $\log(Z(N,x))$ to leading order in $N$.  We did this by fitting
    this data set onto a model function:
    \begin{dmath}
        F(N,x)=c_1(x) N^2 + c_2(x) N \log (N) + c_3(x) N^{} + c_4(x) \log (N) + c_5(x)
    \label{eq:model}
    \end{dmath}
    The form of $F(N,x)$ was decided by examining
    Eq.~\ref{eq:original}. The presence of the restricted double sum
    (Eq.~\ref{eq:N2}) and the leading order behavior of $R_N$ imply
    a $N^2$ dependence of $F(N,x)$, the presence of $N!$ leads to the
    remaining $N$-dependent terms and the constant term is there to
    capture pure temperature dependence of $f^{[\text{max}]}_{N,x}$. We
    studied how these coefficients change with $x$ (or, temperature).
    FIG.~\ref{fig:h1vsx} shows the temperature dependence of $c_1$.
\begin{figure}[htb]
	\centering
		\includegraphics[width=0.80\textwidth]{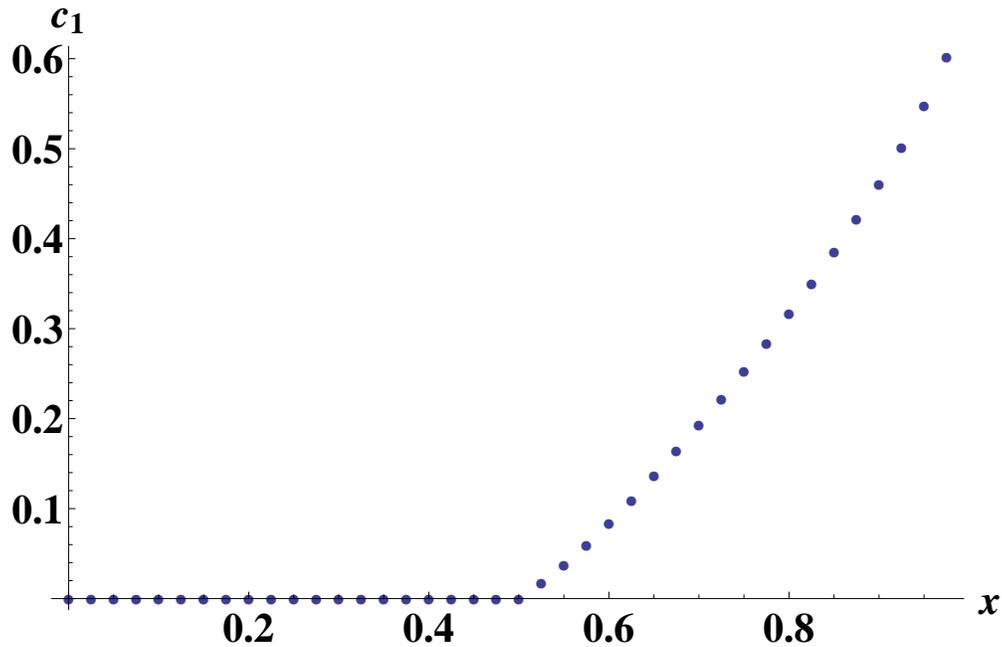}
	\caption{The coefficient of $N^2$ in $\log{Z}$ switches on at
    $x=1/2$.}
	\label{fig:h1vsx}
\end{figure}
    
    \item For each ($N,x$), we used $\{\psi_k\}_{N,x}$ to numerically
    approximate the density function of the global maxima, $\rho(\Psi)$
    (Eq.~\ref{eq:dist}). For $l\in\{1,\cdots,N-1\}$, we obtained
    \begin{dmath}
    \{\rho_l\}_{N,x}=\frac{1}{N-1}\frac{1}{\left\{\psi_{l+1}-\psi_{l}\right\}_{N,x}}
    \end{dmath}  
    where the $1/(N-1)$ is for normalization of the distribution.
\end{enumerate}
\subsection*{Error estimates}
\label{sec:SourcesOfError}
$\widetilde{f}(N,x,\{\theta_k\})$ has multiple maxima of different
orders in the search space $(-\pi,\pi)^{N-1}$. Locating a maximum isn't
as difficult as is ensuring that it is also a global maximum. We
examined four different optimization methods and picked the best one
for obtaining our final data set. A (nonlinear) least-squares fit of
$F(N,x)$ was obtained to the data set. Of the $40$ fits, $38$ had an
Adjusted $R^2$ value of $1$ while two of them had $0.98$ and $0.90$, respectively.
 
\section{Results and Conclusions}
\label{sec:ResultsAndDiscussion}
In FIG.~\ref{fig:max_vs_x} we make an important observation: for
$x<1/2$, $\log(Z)$ is independent of $N$ at leading order. The
$N$-dependence sets in only for $x>1/2$.  For $x>1/2$,
the curves are concave upwards and diverge at $x=1$. Also, as $N$
gets larger, the curve gets steeper. At $N\rightarrow \infty$, the
curve is infinitely steep at $x=1/2$, and we see the Hagedorn
phenomenon.

The most important result of this paper is FIG.~\ref{fig:h1vsx}. This
plot shows the dependence of the leading order coefficient in
Eq.~\ref{eq:model} on temperature. $c_1$ is negligible below $x=1/2$.
In fact, all the coefficients of $N$-dependent terms; i.e.
$c_1,c_2,c_3$ and $c_4$ are negligible below $x=1/2$. We can actually
confirm that there is no $N^2$ term in the expansion of $\log(Z)$
\cite{raha_thorn}. Above $x = 1 / 2$, $c_1$  is positive and increases
monotonically with $x$ (or, temperature). However, it doesn't keep
increasing indefinitely. From Eq.~\ref{eq:x1val} we expect
$\lim_{x \rightarrow 1} c_1 = \log (2)\approx 0.69$. We confirmed this
limiting behavior by conducting a subsequent search with a finer grid
near $x=1$.

FIG.~\ref{fig:h1vsx} clearly demarcates the two temperature regimes we
proposed in the previous section. At lower temperatures $\log (Z)$
has no diverging terms in $N$. Hence, at $N\rightarrow \infty$, it has
a limit point. Above $x=1/2$, the leading order terms diverge as
$N\rightarrow \infty$. It is this difference in behavior that
manifests as the Hagedorn phenomenon as $N\rightarrow \infty$. However,
this difference is present only in the leading order in $N$. As,
has been shown earlier, the finite $N$ partition function; taken in its
entirety, is smooth over $x\in (0,1)$. There is no indication of any
discontinuity or non-differentiability at $x = 1 / 2$. Its only
divergence is at infinite temperature. 

The $N^2$ dependence of $\log{Z}$ is related to the underlying degrees
of freedom in the superstring bit system. It signals the liberation of
the superstring bits from their singlet, polymeric states. In other
words, there is deconfinement in this model at $x=1/2$. The Hagedorn
phenomenon thus has an interpretation as a phase transition between
polymeric, color singlet states and monomeric, color adjoint states.
And in this interpretation, $c_1(x)$ plays the role of the order
parameter. In order to obtain the form of $c_1(x)$ analytically, it is
necessary to find the density of the global maxima,$\{\psi_k\}_{N,x}$,
for $x>1/2$.

\begin{figure}[h]
	\centering
		\includegraphics[width=0.80\textwidth]{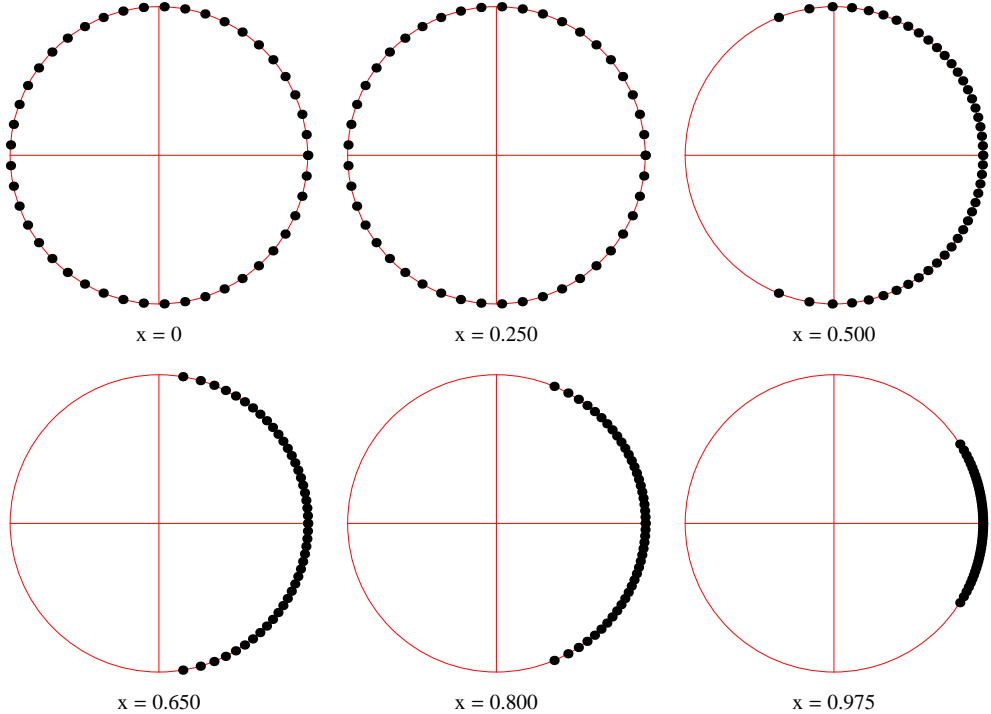}
	\caption{Distribution of the coordinates of the global maxima for $N=45$.}
	\label{fig:polar}
\end{figure}
\begin{figure}[htb]
	\centering
		\includegraphics[width=0.80\textwidth]{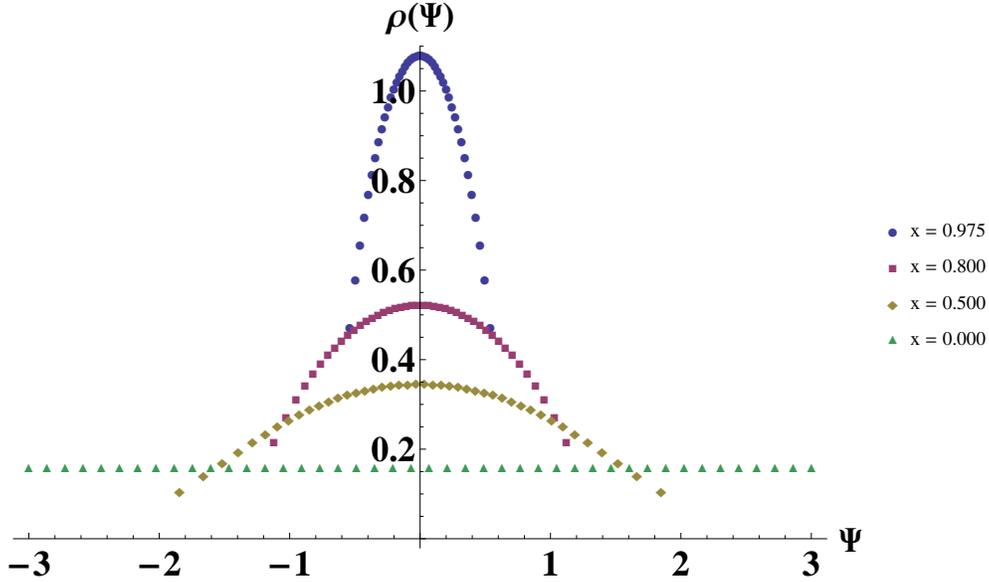}
	\caption{Normalized density function of the {$\psi$}s at different temperatures for $N=45$.}
	\label{fig:density}
\end{figure}

Speaking of $\{\psi_k\}_{N,x}$, this phase transition can also be
detected by examining the distribution of the coordinates of the
global maxima. In FIG.~\ref{fig:polar} we have plotted the
temperature dependence of such distributions for $N=45$. These plots
show a remarkable difference in the distribution of
$\{\psi_k\}_{N,x}$ at different values of temperature. At low
temperatures, the integrand is maximized by those regions in the
domain where the $\{\psi_k\}_{N,x}$ are uniformly distributed in
$(-\pi,\pi)$. It is as if the $\{\psi_k\}_{N,x}$, are
repelling each other. This uniform distribution of of the coordinates
is seen for all $x<1/2$. For $x\geqslant 1/2$, two things happen: the
range of the distribution shrinks and the density peaks at its median.
With increasing temperatures, the $\{\psi_k\}_{N,x}$ begin to attract
each other, until at $x=1$, their distribution becomes a delta
function. At infinite temperature, the integrand is maximized by those
regions where the coordinates have the same value. This change in
distribution of $\{\psi_k\}_{N,x}$ can be rationalized by noting
that $\widehat{f}(0,\{\theta_k\})$ is a function of terms like
$\sin^2\left(\frac{\theta_i-\theta_j}{2}\right)$ and
$\widehat{f}(x\rightarrow 1,\{\theta_k\})$ contains terms like
$\cos^2\left(\frac{\theta_i-\theta_j}{2}\right)$. However, it is
interesting that this transition of the distribution doesn't begin
until $x=1/2$. One can construct the normalized density function from
the $\{\psi_k\}_{N,x}$. FIG.~\ref{fig:density} shows these density
plots at different temperatures for $N=45$. Again, till $x=1/2$, the
density function is a constant, with $\rho\approx1/(2\pi)$. Above
$x=1/2$, there exists a $\Psi_{[\text{max}]}< \pi$ such that
$\rho\left(\pm\Psi_{[\text{max}]}\right)=0$ (Eq.~\ref{eq:maxro}). Such
a cut-off for the density function has featured in previous studies of
matrix models, e.g. the Hermitian matrix model \cite{brezin_planar_1978}.

The change in the distribution of $\{\psi_k\}$ in our superstring
bit model, bears a resemblance to the well-known phase transition in
the unitary matrix model \cite{gross_possible_1980}. While the latter
has a coupling constant $g$, the parameter in our model is $x$, or the
temperature. Still, there is a similarity in the transformation of the
density functions: from low to high temperature in superstring bit
model and from small to large coupling in the unitary matrix model. 
Recently, it was pointed out to us that Aharony et al
 \footnote{Sundborg \cite{sundborg_hagedorn_2000} had, in turn,
  already derived most of the results of
   \cite{aharony_hagedorn/deconfinement_2004} for N=4 SYM theory on $S^3$.} obtained similar results for free $U(N)$ Yang-Mills theory with
   adjoint matter on $S^3\times\mathbb{R}$ \cite{aharony_hagedorn/deconfinement_2004}. Both Sundborg
      \cite{sundborg_hagedorn_2000} and \cite{aharony_hagedorn/deconfinement_2004}
      have obtained similar supersymmetric partition functions in the
      limit of large $N$. However, unlike in these models, an exact
      analytic expression of the high temperature density
       function still evades us.

\section*{Acknowledgments}
I thank Charles Thorn for his insights and guidance in this project. I
thank Thomas Curtright for pointing us to McKay's papers
\cite{mckay_applications_1983,mckay_asymptotic_1990}. I thank David McGady for pointing us to existing literature on Hagedorn phenomena, esp.  \cite{aharony_hagedorn/deconfinement_2004,sundborg_hagedorn_2000}. This work was supported in part by the Department of Energy under Grant No. DE-SC0010296.

\appendix*
\section{Gaussian approximation at infinite temperature}
\label{sec:deriv}

Let us Taylor-expand $\widehat{f}$ from Eq.~\ref{eq:gauss} in \{$\theta_k$\}s about a global maximum at \{$\psi_k$\}:
\begin{dmath}
\widehat{f} \left(x,\{\theta_k\}\right) = \widehat{f} \left(x,\{\psi_k\}\right)
 + \frac{1}{2} \sum_{r, s} (\theta_r - \psi_r) \frac{d^2}{d \theta_r d \theta_s} \widehat{f}\left(x,\{\psi_k\}\right)  (\theta_s   - \psi_s) + \cdots
\end{dmath}
\begin{dmath}
\implies \int_{-\pi}^{\pi} \exp \widehat{f} \left(x,\{\theta_k\}\right)
   \prod_{1 \leqslant l \leqslant N} d \theta_l
\approx  \mathcal{J}(N,x)\exp\left\{\widehat{f} \left(x,\{\psi_k\}\right)\right\}
\int_{-\infty}^\infty \exp\left\{ \frac{1}{2} \sum_{r, s} (\theta_r - \psi_r) \frac{d^2 \widehat{f}\left(x,\{\psi_k\}\right)}{d \theta_r d \theta_s} (\theta_s   - \psi_s) \right\} \prod_{1 \leqslant l \leqslant N} d \theta_l
=\mathcal{J}(N,x)\exp\left\{\widehat{f} \left(x,\{\psi_k\}\right)\right\} \sqrt{\frac{(2\pi)^N}{\text{det}(H)}}
\end{dmath}

where $\mathcal{J}(N,x)$ is the number of global maxima in $(-\pi,\pi)^N$ and \[H_{rs}= -\frac{d^2}{d \theta_r d \theta_s} \widehat{f}\left(x,\{\psi_k\}\right)\]
As the integrand is translation invariant in the $\theta$s,
$\mathcal{J}(N,x)$ is $\infty$. As we shall see later, the correct
approximation involves applying steepest descent after removing the
zero mode integral. Once the zero mode is removed, $\mathcal{J}(N,x)$
is the number of global maxima in $(-\pi,\pi)^{N-1}$. $\mathcal{J}(N,x)$
in general has a factor of $(N-1)!$ coming from the symmetry of
the integrand in the $\theta$s. 

At infinite temperature ($x=1$) there is only one global maximum:
$\psi_1 = \psi_2 = \cdots =\psi_N$. Hence, $\mathcal{J}(N,1)=1$ and
from Eq.~\ref{eq:high}
\begin{widetext}
\begin{dmath}
  Z(N,1) \approx\left( \frac{2}{1 - x}  \right)^{N - 1} \frac{2^{N(N-1)} }{N! (2  
\pi)^N} \int^\infty_{-\infty}  \exp\left\{-\frac{1}{2}\sum_{r , s}\underbrace{\frac{N\delta_{r s}-1}{2}}_{=H_{rs}} \theta_r \theta_s \right\} \prod^N_{k = 1} d \theta_k
\end{dmath}
\end{widetext}

Here we can see that the formula for the Gaussian integral yields
$\infty$. This is because we blindly replaced every one dimensional
integration $\int_{-\pi}^\pi {d\theta_i \ \cdots}$ with
$\int_{-\infty}^\infty {d\theta_i\ \cdots}$ when we took the Gaussian
approximation. Hence, instead of obtaining a factor of $2\pi$ from the
zero mode, we get $\infty$. The correct formula is
\[2\pi \sqrt{\frac{(2\pi)^{N-1}}{\text{det}(H')}}\]
where $H'_{rs}=H_{rs}\ \forall \{r,s\} \ \in \{1,2,\cdots,N-1\}$. $H'$
is the $(N-1)\times(N-1)$ matrix one gets after truncating the
$N^{\text{th}}$ row and $N^{\text{th}}$ column of $H$. Evaluating
$\text{det}(H')$ we get $\frac{N^{N-2}}{2^{N-1}}$. Putting everything
back into the earlier equation we get
\begin{dmath}
  Z \approx\left( \frac{2}{1 - x}  \right)^{N - 1} \frac{2^{N(N-1)} }{N! (2  \pi)^N} 2\pi \sqrt{\frac{2^{N-1}(2\pi)^{N-1}}{N^{N-2}}}=\left( \frac{2}{1 - x}  \right)^{N - 1}\frac{1}{N!} \frac{2^{N(N-1)} }{\pi^{(N-1)/2}N^{N/2-1}}
\end{dmath}
We can compare this to Eq.~\ref{eq:high} to infer:
\begin{dmath}R_N=\sum_\Gamma{\text{Res}  \left(  \frac{\prod_{1 \leqslant i < j \leqslant N}  (z_i +  z_j)^2}{\prod_{1 \leqslant k \leqslant N}  (z_k)^N} \right)}\approx \frac{2^{N(N-1)} }{\pi^{(N-1)/2}N^{N/2-1}}
\label{eq:last}
\end{dmath}

While searching through mathematics literature, we came to know that
$R_N$ as defined above, also counts the number of Eulerian digraphs
with $N$ nodes \cite{mckay_applications_1983}. Also, the author lists
exact values of $R_N$ till $N=16$, some of which we have used in
FIG.~\ref{fig:hightemp}. A follow up on our search revealed that the
asymptotic expression in Eq.~\ref{eq:last} has already been computed in
\cite{mckay_asymptotic_1990}. It turns out that multiplying our
Gaussian result with an extra factor of $e^{-1/4}$ is a more accurate
approximation in leading order.

%\bibliography{preprint,ref}
\input{preprint.bbl}

\end{document}

%% file: preprint.bbl

%% file: preprint.bbl
\begin{thebibliography}{20}%
\makeatletter
\providecommand \@ifxundefined [1]{%
 \@ifx{#1\undefined}
}%
\providecommand \@ifnum [1]{%
 \ifnum #1\expandafter \@firstoftwo
 \else \expandafter \@secondoftwo
 \fi
}%
\providecommand \@ifx [1]{%
 \ifx #1\expandafter \@firstoftwo
 \else \expandafter \@secondoftwo
 \fi
}%
\providecommand \natexlab [1]{#1}%
\providecommand \enquote  [1]{``#1''}%
\providecommand \bibnamefont  [1]{#1}%
\providecommand \bibfnamefont [1]{#1}%
\providecommand \citenamefont [1]{#1}%
\providecommand \href@noop [0]{\@secondoftwo}%
\providecommand \href [0]{\begingroup \@sanitize@url \@href}%
\providecommand \@href[1]{\@@startlink{#1}\@@href}%
\providecommand \@@href[1]{\endgroup#1\@@endlink}%
\providecommand \@sanitize@url [0]{\catcode `\\12\catcode `\$12\catcode
  `\&12\catcode `\#12\catcode `\^12\catcode `\_12\catcode `\%12\relax}%
\providecommand \@@startlink[1]{}%
\providecommand \@@endlink[0]{}%
\providecommand \url  [0]{\begingroup\@sanitize@url \@url }%
\providecommand \@url [1]{\endgroup\@href {#1}{\urlprefix }}%
\providecommand \urlprefix  [0]{URL }%
\providecommand \Eprint [0]{\href }%
\providecommand \doibase [0]{http://dx.doi.org/}%
\providecommand \selectlanguage [0]{\@gobble}%
\providecommand \bibinfo  [0]{\@secondoftwo}%
\providecommand \bibfield  [0]{\@secondoftwo}%
\providecommand \translation [1]{[#1]}%
\providecommand \BibitemOpen [0]{}%
\providecommand \bibitemStop [0]{}%
\providecommand \bibitemNoStop [0]{.\EOS\space}%
\providecommand \EOS [0]{\spacefactor3000\relax}%
\providecommand \BibitemShut  [1]{\csname bibitem#1\endcsname}%
\let\auto@bib@innerbib\@empty
%</preamble>
\bibitem [{\citenamefont {Giles}\ and\ \citenamefont
  {Thorn}(1977)}]{giles_lattice_1977}%
  \BibitemOpen
  \bibfield  {author} {\bibinfo {author} {\bibfnamefont {Roscoe}\ \bibnamefont
  {Giles}}\ and\ \bibinfo {author} {\bibfnamefont {Charles~B.}\ \bibnamefont
  {Thorn}},\ }\bibfield  {title} {\enquote {\bibinfo {title} {Lattice approach
  to string theory},}\ }\href {\doibase 10.1103/PhysRevD.16.366} {\bibfield
  {journal} {\bibinfo  {journal} {Physical Review D}\ }\textbf {\bibinfo
  {volume} {16}},\ \bibinfo {pages} {366--386} (\bibinfo {year}
  {1977})}\BibitemShut {NoStop}%
\bibitem [{\citenamefont {Hooft}(1974)}]{hooft_planar_1974}%
  \BibitemOpen
  \bibfield  {author} {\bibinfo {author} {\bibfnamefont {G.~'t}\ \bibnamefont
  {Hooft}},\ }\bibfield  {title} {\enquote {\bibinfo {title} {A planar diagram
  theory for strong interactions},}\ }\href {\doibase
  10.1016/0550-3213(74)90154-0} {\bibfield  {journal} {\bibinfo  {journal}
  {Nuclear Physics B}\ }\textbf {\bibinfo {volume} {72}},\ \bibinfo {pages}
  {461--473} (\bibinfo {year} {1974})}\BibitemShut {NoStop}%
\bibitem [{\citenamefont {Thorn}(1994)}]{thorn_reformulating_1994}%
  \BibitemOpen
  \bibfield  {author} {\bibinfo {author} {\bibfnamefont {Charles~B.}\
  \bibnamefont {Thorn}},\ }\bibfield  {title} {\enquote {\bibinfo {title}
  {Reformulating {String} {Theory} with the \$1/{N}\$ {Expansion}},}\ }\href
  {http://arxiv.org/abs/hep-th/9405069} {\bibfield  {journal} {\bibinfo
  {journal} {arXiv:hep-th/9405069}\ } (\bibinfo {year} {1994})},\ \bibinfo
  {note} {arXiv: hep-th/9405069}\BibitemShut {NoStop}%
\bibitem [{\citenamefont {Bergman}\ and\ \citenamefont
  {Thorn}(1995)}]{bergman_string_1995}%
  \BibitemOpen
  \bibfield  {author} {\bibinfo {author} {\bibfnamefont {Oren}\ \bibnamefont
  {Bergman}}\ and\ \bibinfo {author} {\bibfnamefont {Charles~B.}\ \bibnamefont
  {Thorn}},\ }\bibfield  {title} {\enquote {\bibinfo {title} {String bit models
  for superstring},}\ }\href {\doibase 10.1103/PhysRevD.52.5980} {\bibfield
  {journal} {\bibinfo  {journal} {Physical Review D}\ }\textbf {\bibinfo
  {volume} {52}},\ \bibinfo {pages} {5980--5996} (\bibinfo {year}
  {1995})}\BibitemShut {NoStop}%
\bibitem [{\citenamefont {Thorn}(2014)}]{thorn_space_2014}%
  \BibitemOpen
  \bibfield  {author} {\bibinfo {author} {\bibfnamefont {Charles~B.}\
  \bibnamefont {Thorn}},\ }\bibfield  {title} {\enquote {\bibinfo {title}
  {Space from string bits},}\ }\href {\doibase 10.1007/JHEP11(2014)110}
  {\bibfield  {journal} {\bibinfo  {journal} {Journal of High Energy Physics}\
  }\textbf {\bibinfo {volume} {2014}},\ \bibinfo {pages} {1--24} (\bibinfo
  {year} {2014})}\BibitemShut {NoStop}%
\bibitem [{\citenamefont {Hagedorn}(1965)}]{hagedorn_statistical_1965}%
  \BibitemOpen
  \bibfield  {author} {\bibinfo {author} {\bibfnamefont {R.}~\bibnamefont
  {Hagedorn}},\ }\bibfield  {title} {\enquote {\bibinfo {title} {Statistical
  thermodynamics of strong interactions at high-energies},}\ }\href@noop {}
  {\bibfield  {journal} {\bibinfo  {journal} {Nuovo Cim.Suppl.}\ }\textbf
  {\bibinfo {volume} {3}},\ \bibinfo {pages} {147--186} (\bibinfo {year}
  {1965})}\BibitemShut {NoStop}%
\bibitem [{\citenamefont {Fubini}\ and\ \citenamefont
  {Veneziano}(1969)}]{fubini_level_1969}%
  \BibitemOpen
  \bibfield  {author} {\bibinfo {author} {\bibfnamefont {S.}~\bibnamefont
  {Fubini}}\ and\ \bibinfo {author} {\bibfnamefont {G.}~\bibnamefont
  {Veneziano}},\ }\bibfield  {title} {\enquote {\bibinfo {title} {Level
  structure of dual-resonance models},}\ }\href {\doibase 10.1007/BF02758835}
  {\bibfield  {journal} {\bibinfo  {journal} {Il Nuovo Cimento A (1971-1996)}\
  }\textbf {\bibinfo {volume} {64}},\ \bibinfo {pages} {811--840} (\bibinfo
  {year} {1969})}\BibitemShut {NoStop}%
\bibitem [{\citenamefont {Atick}\ and\ \citenamefont
  {Witten}(1988)}]{atick_hagedorn_1988}%
  \BibitemOpen
  \bibfield  {author} {\bibinfo {author} {\bibfnamefont {Joseph~J.}\
  \bibnamefont {Atick}}\ and\ \bibinfo {author} {\bibfnamefont {Edward}\
  \bibnamefont {Witten}},\ }\bibfield  {title} {\enquote {\bibinfo {title} {The
  {Hagedorn} transition and the number of degrees of freedom of string
  theory},}\ }\href {\doibase 10.1016/0550-3213(88)90151-4} {\bibfield
  {journal} {\bibinfo  {journal} {Nuclear Physics B}\ }\textbf {\bibinfo
  {volume} {310}},\ \bibinfo {pages} {291--334} (\bibinfo {year}
  {1988})}\BibitemShut {NoStop}%
\bibitem [{\citenamefont {Thorn}(2015)}]{thorn_string_2015}%
  \BibitemOpen
  \bibfield  {author} {\bibinfo {author} {\bibfnamefont {Charles~B.}\
  \bibnamefont {Thorn}},\ }\bibfield  {title} {\enquote {\bibinfo {title}
  {String bits at finite temperature and the {Hagedorn} phase},}\ }\href
  {\doibase 10.1103/PhysRevD.92.066007} {\bibfield  {journal} {\bibinfo
  {journal} {Physical Review D}\ }\textbf {\bibinfo {volume} {92}} (\bibinfo
  {year} {2015}),\ 10.1103/PhysRevD.92.066007}\BibitemShut {NoStop}%
\bibitem [{\citenamefont {Sun}\ and\ \citenamefont
  {Thorn}(2014)}]{sun_stable_2014}%
  \BibitemOpen
  \bibfield  {author} {\bibinfo {author} {\bibfnamefont {Songge}\ \bibnamefont
  {Sun}}\ and\ \bibinfo {author} {\bibfnamefont {Charles~B.}\ \bibnamefont
  {Thorn}},\ }\bibfield  {title} {\enquote {\bibinfo {title} {Stable string bit
  models},}\ }\href {\doibase 10.1103/PhysRevD.89.105002} {\bibfield  {journal}
  {\bibinfo  {journal} {Physical Review D}\ }\textbf {\bibinfo {volume} {89}},\
  \bibinfo {pages} {105002} (\bibinfo {year} {2014})}\BibitemShut {NoStop}%
\bibitem [{\citenamefont {Chen}\ and\ \citenamefont
  {Sun}(2016)}]{chen_numerical_2016}%
  \BibitemOpen
  \bibfield  {author} {\bibinfo {author} {\bibfnamefont {Gaoli}\ \bibnamefont
  {Chen}}\ and\ \bibinfo {author} {\bibfnamefont {Songge}\ \bibnamefont
  {Sun}},\ }\bibfield  {title} {\enquote {\bibinfo {title} {Numerical study of
  the simplest string bit model},}\ }\href {\doibase
  10.1103/PhysRevD.93.106004} {\bibfield  {journal} {\bibinfo  {journal}
  {Physical Review D}\ }\textbf {\bibinfo {volume} {93}},\ \bibinfo {pages}
  {106004} (\bibinfo {year} {2016})}\BibitemShut {NoStop}%
\bibitem [{\citenamefont {Aharony}\ \emph {et~al.}(2004)\citenamefont
  {Aharony}, \citenamefont {Marsano}, \citenamefont {Minwalla}, \citenamefont
  {Papadodimas},\ and\ \citenamefont
  {Van~Raamsdonk}}]{aharony_hagedorn/deconfinement_2004}%
  \BibitemOpen
  \bibfield  {author} {\bibinfo {author} {\bibfnamefont {Ofer}\ \bibnamefont
  {Aharony}}, \bibinfo {author} {\bibfnamefont {Joseph}\ \bibnamefont
  {Marsano}}, \bibinfo {author} {\bibfnamefont {Shiraz}\ \bibnamefont
  {Minwalla}}, \bibinfo {author} {\bibfnamefont {Kyriakos}\ \bibnamefont
  {Papadodimas}}, \ and\ \bibinfo {author} {\bibfnamefont {Mark}\ \bibnamefont
  {Van~Raamsdonk}},\ }\bibfield  {title} {\enquote {\bibinfo {title} {The
  {Hagedorn}/{Deconfinement} {Phase} {Transition} in {Weakly} {Coupled} {Large}
  {N} {Gauge} {Theories}},}\ }\href {\doibase 10.4310/ATMP.2004.v8.n4.a1}
  {\bibfield  {journal} {\bibinfo  {journal} {Advances in Theoretical and
  Mathematical Physics}\ }\textbf {\bibinfo {volume} {8}},\ \bibinfo {pages}
  {603--696} (\bibinfo {year} {2004})}\BibitemShut {NoStop}%
\bibitem [{\citenamefont {Sundborg}(2000)}]{sundborg_hagedorn_2000}%
  \BibitemOpen
  \bibfield  {author} {\bibinfo {author} {\bibfnamefont {Bo}~\bibnamefont
  {Sundborg}},\ }\bibfield  {title} {\enquote {\bibinfo {title} {The {Hagedorn}
  transition, deconfinement and {N}=4 {SYM} theory},}\ }\href {\doibase
  10.1016/S0550-3213(00)00044-4} {\bibfield  {journal} {\bibinfo  {journal}
  {Nuclear Physics B}\ }\textbf {\bibinfo {volume} {573}},\ \bibinfo {pages}
  {349--363} (\bibinfo {year} {2000})}\BibitemShut {NoStop}%
\bibitem [{\citenamefont {Curtright}\ \emph {et~al.}(2017)\citenamefont
  {Curtright}, \citenamefont {Raha},\ and\ \citenamefont {Thorn}}]{raha_thorn}%
  \BibitemOpen
  \bibfield  {author} {\bibinfo {author} {\bibfnamefont {Thomas~L.}\
  \bibnamefont {Curtright}}, \bibinfo {author} {\bibfnamefont {Sourav}\
  \bibnamefont {Raha}}, \ and\ \bibinfo {author} {\bibfnamefont {Charles~B.}\
  \bibnamefont {Thorn}},\ }\bibfield  {title} {\enquote {\bibinfo {title}
  {Color {Characters} for {White} {Hot} {String} {Bits}},}\ }\href
  {http://arxiv.org/abs/1708.03342} {\bibfield  {journal} {\bibinfo  {journal}
  {arXiv:1708.03342 [hep-th]}\ } (\bibinfo {year} {2017})},\ \bibinfo {note}
  {arXiv: 1708.03342}\BibitemShut {NoStop}%
\bibitem [{\citenamefont {Weyl}(1946)}]{weyl}%
  \BibitemOpen
  \bibfield  {author} {\bibinfo {author} {\bibfnamefont {Hermann}\ \bibnamefont
  {Weyl}},\ }\href@noop {} {\emph {\bibinfo {title} {The Classical Groups,
  Their Invariants and Representations}}}\ (\bibinfo  {publisher} {Princeton
  University Press},\ \bibinfo {address} {Princeton, N.J.},\ \bibinfo {year}
  {1946})\BibitemShut {NoStop}%
\bibitem [{\citenamefont {Curtright}\ and\ \citenamefont
  {Thorn}(1986)}]{curtright_symmetry_1986}%
  \BibitemOpen
  \bibfield  {author} {\bibinfo {author} {\bibfnamefont {Thomas~L.}\
  \bibnamefont {Curtright}}\ and\ \bibinfo {author} {\bibfnamefont
  {Charles~B.}\ \bibnamefont {Thorn}},\ }\bibfield  {title} {\enquote {\bibinfo
  {title} {Symmetry patterns in the mass spectra of dual string models},}\
  }\href {\doibase 10.1016/0550-3213(86)90525-0} {\bibfield  {journal}
  {\bibinfo  {journal} {Nuclear Physics B}\ }\textbf {\bibinfo {volume}
  {274}},\ \bibinfo {pages} {520--558} (\bibinfo {year} {1986})}\BibitemShut
  {NoStop}%
\bibitem [{\citenamefont {McKay}(1983)}]{mckay_applications_1983}%
  \BibitemOpen
  \bibfield  {author} {\bibinfo {author} {\bibfnamefont {Brendan~D.}\
  \bibnamefont {McKay}},\ }\bibfield  {title} {\enquote {\bibinfo {title}
  {Applications of a technique for labelled enumeration},}\ }\href
  {http://cs.anu.edu.au/~./bdm/papers/LabelledEnumeration.pdf} {\bibfield
  {journal} {\bibinfo  {journal} {Congressus Numerantium}\ }\textbf {\bibinfo
  {volume} {40}},\ \bibinfo {pages} {207--221} (\bibinfo {year}
  {1983})}\BibitemShut {NoStop}%
\bibitem [{\citenamefont {Br\'ezin}\ \emph {et~al.}(1978)\citenamefont
  {Br\'ezin}, \citenamefont {Itzykson}, \citenamefont {Parisi},\ and\
  \citenamefont {Zuber}}]{brezin_planar_1978}%
  \BibitemOpen
  \bibfield  {author} {\bibinfo {author} {\bibfnamefont {E.}~\bibnamefont
  {Br\'ezin}}, \bibinfo {author} {\bibfnamefont {C.}~\bibnamefont {Itzykson}},
  \bibinfo {author} {\bibfnamefont {G.}~\bibnamefont {Parisi}}, \ and\ \bibinfo
  {author} {\bibfnamefont {J.~B.}\ \bibnamefont {Zuber}},\ }\bibfield  {title}
  {\enquote {\bibinfo {title} {Planar diagrams},}\ }\href {\doibase
  10.1007/BF01614153} {\bibfield  {journal} {\bibinfo  {journal}
  {Communications in Mathematical Physics}\ }\textbf {\bibinfo {volume} {59}},\
  \bibinfo {pages} {35--51} (\bibinfo {year} {1978})}\BibitemShut {NoStop}%
\bibitem [{\citenamefont {Gross}\ and\ \citenamefont
  {Witten}(1980)}]{gross_possible_1980}%
  \BibitemOpen
  \bibfield  {author} {\bibinfo {author} {\bibfnamefont {David~J.}\
  \bibnamefont {Gross}}\ and\ \bibinfo {author} {\bibfnamefont {Edward}\
  \bibnamefont {Witten}},\ }\bibfield  {title} {\enquote {\bibinfo {title}
  {Possible third-order phase transition in the large-\${N}\$ lattice gauge
  theory},}\ }\href {\doibase 10.1103/PhysRevD.21.446} {\bibfield  {journal}
  {\bibinfo  {journal} {Physical Review D}\ }\textbf {\bibinfo {volume} {21}},\
  \bibinfo {pages} {446--453} (\bibinfo {year} {1980})}\BibitemShut {NoStop}%
\bibitem [{\citenamefont {McKay}(1990)}]{mckay_asymptotic_1990}%
  \BibitemOpen
  \bibfield  {author} {\bibinfo {author} {\bibfnamefont {Brendan~D.}\
  \bibnamefont {McKay}},\ }\bibfield  {title} {\enquote {\bibinfo {title} {The
  asymptotic numbers of regular tournaments, {Eulerian} digraphs and {Eulerian}
  oriented graphs},}\ }\href {\doibase 10.1007/BF02128671} {\bibfield
  {journal} {\bibinfo  {journal} {Combinatorica}\ }\textbf {\bibinfo {volume}
  {10}},\ \bibinfo {pages} {367--377} (\bibinfo {year} {1990})}\BibitemShut
  {NoStop}%
\end{thebibliography}
